\newcount\Comments  
\documentclass[preprint,12pt]{elsarticle}

\usepackage{graphicx}
\usepackage{amssymb}
\usepackage{amsmath}
\usepackage[]{algorithm2e}
\usepackage{mathrsfs}
\DeclareMathOperator{\EX}{\mathbb{E}}

\usepackage{color}
\definecolor{darkgreen}{rgb}{0,0.5,0}
\definecolor{purple}{rgb}{1,0,1}
\newcommand{\kibitz}[2]{\ifnum\Comments=1\textcolor{#1}{#2}\fi}

\journal{Journal of Theoretical Biology}

\begin{document}

\begin{frontmatter}


\title{On the probability of strain invasion in endemic settings: accounting for individual heterogeneity and control in multi-strain dynamics}



\author[1,*]{Michael T. Meehan}

\author[2,*]{Robert C. Cope}

\author[1]{Emma S. McBryde}

\address[1]{James Cook University, Australian Institute of Tropical Health and Medicine, Townsville, Australia}

\address[2]{The University of Adelaide, School of Mathematical Sciences, Adelaide, Australia}

\address[*]{These authors contributed equally to this work}

\begin{abstract}

The rise of antimicrobial drug resistance is an imminent threat to global health that has warranted, and duly received, considerable attention within the medical, microbiological and modelling communities. 
Outbreaks of drug-resistant pathogens are ignited by the emergence and transmission of mutant variants descended from wild-type strains circulating in the community. In this work we investigate the stochastic dynamics of the emergence of a novel disease strain, introduced into a population in which it must compete with an existing endemic strain. In analogy with past work on single-strain epidemic outbreaks, we apply a branching process approximation to calculate the probability that the new strain becomes established.
As expected, a critical determinant of the survival prospects of any invading strain is the magnitude of its reproduction number relative to that of the background endemic strain. Whilst in most circumstances this ratio must exceed unity in order for invasion to be viable, we show that differential control scenarios can lead to less-fit novel strains invading populations hosting a fitter endemic one. This analysis and the accompanying findings will inform our understanding of the mechanisms that have led to past instances of successful strain invasion, and provide valuable lessons for thwarting future drug-resistant strain incursions. 
\end{abstract}


\end{frontmatter}


\section{Introduction}
\label{intro}

Antimicrobial drug resistance (AMR) currently presents one of the most significant challenges to public health worldwide~\cite{levy2004antibacterial}. The incidence of cases exhibiting resistance to both first- and second-line antibiotic treatments continues to rise with endemic levels already witnessed in several regions~\cite{mckenna2013antibiotic,laxminarayan2013antibiotic}. Exacerbating the problem is the growing misuse and overuse of antimicrobial agents which, rather than suppressing pathogen strain diversity, often acts to accelerate the spread and diversification of drug-resistant variants~\cite{barbosa2000impact,Neu1992}. Alarmingly, in many settings, direct transmission of these mutant pathogens is believed to have become the primary source of incident cases~\cite{luciani2009epidemiological,world2018global} suggesting that intervention strategies targeting the transmission and treatment of drug-sensitive strains alone will likely fail. Even with recent increases in funding to combat the rise of antimicrobial resistance~\cite{simpkin2017incentivizing}, the ongoing arms race between pathogen evolution and the development of novel antibiotics is still being won by the former. Consequently, we are now faced with a rapidly diminishing arsenal of effective therapies --- one which may soon prove inadequate~\cite{hancock2007end}.

The rise of antimicrobial resistance is driven by the evolution of infectious pathogens and the emergence of mutant variants which, by its very nature, is a random process --- both at the within-host and population levels. Not only do mutant strains arise as a result of random microbiological processes~\cite{alekshun2007molecular}, but phenotypically distinct variants initially appear in small numbers of hosts and must therefore escape a treacherous stochastic regime (i.e., avoid extinction) in order to establish themselves within the community. Often furthering this challenge is the presence of a resident endemic (possibly ancestral) strain that deprives the newly-emergent mutant of susceptible hosts (e.g., through cross-immunity). In spite of these challenges, countless mutant pathogens --- often exhibiting varying levels of drug-resistance --- have flourished and been able to firmly entrench themselves among host populations~\cite{Chang2015origin}.

Previous modelling investigations of pathogen ecology and diversity have often assumed that the multi-strain dynamics are well established
~\cite{Colijn2010mechanism, Cobey2017host,Lehtinan2017evolutionary, Blanquart2017evolutionary, meehan2018};
whilst those studies that have specifically modelled the emergence of drug-resistant variants, have predominantly done so deterministically
(see e.g.,~\cite{Cohen2004modeling, Blower2004modeling,tanaka2006detecting,Dagata2009modeling}). The goal of this paper is to analyze the early-time dynamics of novel strain emergence in a stochastic setting whilst accounting for the background population dynamics associated with endemic infection. In particular, we investigate the conditions under which invasion by a novel strain is possible and calculate the probability of this event occurring, depending on the relative fitness (i.e., reproductive capacity) of the two strains, and the relevant characteristics of the host population (i.e., the heterogeneity in individual reproductive capacity). We then analyze the effects of control strategies and timing on the emergence of novel strains and discuss the implications of differential control efficacy when the invading strain boasts some level of resistance to treatment. The results of this investigation will extend analogous findings in single-strain settings and elucidate the consequences of different control mechanisms/timings in multi-strain systems --- particularly the impact of differential control. Moreover the analytical framework that we develop will provide a flexible and robust theoretical basis for future modelling work.

\section{Background: invasion in infection-na{\"i}ve populations}
\label{sec:bg}

Before turning to the multi-strain scenario, we first review the analogous computation in an infection-na{\" i}ve setting --- for which several classical results are available~\cite{griffiths1973multivariate,britton2010stochastic,diekmann2012mathematical}: the probability of an epidemic outbreak when an infectious disease is introduced into an entirely susceptible host population.

To begin, when the susceptible population is sufficiently large, it is common to approximate early epidemic behavior by a branching process~\cite{griffiths1973multivariate,britton2010stochastic,diekmann2012mathematical}, as infected individuals are likely to have contacts only with susceptible individuals. In this case the depletion of susceptible individuals due to infection is negligible such that (in an infection-na{\" i}ve population) the susceptible fraction remains approximately equal to one. Thus, to determine the epidemic trajectory we need only track the number of infected individuals and the rate at which they produce secondary cases, or offspring. This framework allows for tractable computations of the probability an outbreak becomes extinct in its early stages, using branching process theory.

When our goal is solely to determine the probability of extinction or establishment of the outbreak, we need only consider the infection process in terms of discrete, non-overlapping generations. If the duration of infectiousness is constant, the number of secondary cases, or offspring, $Z$, generated by each member of the infectious population during their infectious lifetime (i.e., one generation) is a Poisson random variable with rate parameter $R_0$, where $R_0$ is the basic reproduction number: $Z \sim \mathrm{Poisson}(R_0)$. In this case, the probability that an epidemic outbreak, that begins with a single infectious individual, goes extinct, $q$, is given by the smallest root of the equation (see e.g.,~\cite{diekmann2012mathematical}):
\begin{equation}
    q = \mathrm{e}^{-R_0(1-q)}.\label{eq:extinction_special}
\end{equation}
For $R_0 \leq 1$ the unique root of equation~\eqref{eq:extinction_special} is $q = 1$ and extinction is guaranteed. For $R_0 > 1$ a second solution $q < 1$ emerges and an epidemic outbreak is possible. This outcome becomes increasingly likely as $R_0$ increases. 

A significant recent contribution to this theory is the work by Lloyd-Smith et al.~\cite{superspreaders} in which the authors highlighted the importance of heterogeneity in individual infectiousness among members of the host population and its role in determining the fate of an infectious disease outbreak. By analyzing secondary case counts from several historical infectious disease outbreaks, including the 2003 outbreak of severe acute respiratory syndrome (SARS) in Singapore, the authors demonstrated  that in order to accurately reconstruct observed epidemics, models must account for individual heterogeneity in reproductive potential. In particular, they found that observed secondary case counts follow highly skewed distributions where outbreaks are driven by a small subset of so-called `super-spreaders': individuals that generate a disproportionately large number of secondary cases, and that the large majority of infected individuals lead to no, or few, secondary infections. As a consequence of this disparity, the probability of extinction increases and epidemics become rarer (but more explosive), as heterogeneity increases. Importantly, their findings show that previous analyses that rely solely on population mean parameters --- such as the basic reproduction number $R_0$ --- to model outbreak dynamics inadequately capture the true stochastic nature of disease spread, because the prospects of an epidemic outbreak depend critically on the distribution of infection potential within the host population.

To model the infectious cohort as a heterogeneous mixture of infected individuals, the authors in~\cite{superspreaders} assigned each member of the infectious population a random reproductive capacity $\nu$ drawn from a Gamma probability distribution with population mean $R_0 = \EX(\nu)$ and dispersion (i.e. shape) parameter $k$. As such, the probability density function for $\nu$ was defined as
\begin{equation}
    p_\nu(x) = \frac{1}{\Gamma(k)}\,\frac{k}{R_0}\left(\frac{kx}{R_0}\right)^{k-1}\mathrm{e}^{-kx/R_0}.\nonumber
\end{equation}
Here, smaller values of $k$ indicate higher levels of heterogeneity in the infected population and a greater propensity for super-spreading behaviour.
This variation in reproductive capacity can be interpreted as representing some combination of, for example, variation in duration of infectiousness, differences in social contacts or behaviour between individuals, or higher transmissibility among children as compared to adults. Being able to represent this heterogeneity without needing to explicitly construct social contact networks or multi-class processes provides a flexible and tractable means of exploring a range of scenarios. More importantly, the Gamma distribution specifically, in conjunction with the branching process approximation, was shown in~\cite{superspreaders} to accurately reproduce the secondary case counts of several observed epidemic outbreaks. For these reasons we adopt this modelling approach in our investigation.

When the reproductive potential $\nu$ is Gamma distributed, the number of offspring $Z$ follows a Negative Binomial distribution with parameters $p$ and $k$: $Z\sim \mathrm{NegBin}(p, k)$, with $p = (1 + \frac{R_0}{k})^{-1}$. The extinction equation~\eqref{eq:extinction_special} generalizes to~\cite{superspreaders}
\begin{equation}
    q = \left(1 + \frac{R_0}{k}(1-q)\right)^{-k}.\label{eq:extinction}
\end{equation}
In the limit $k\rightarrow\infty$, that is, for a homogeneous population with $\nu=R_0$, this equation reduces to the familiar form~\eqref{eq:extinction_special}, and for $k=1$ (i.e., the typical continuous-time Markov chain with Exponentially-distributed infectious periods and inter-arrival times between infection events), we have the simple expression
\begin{equation}
    q = \frac{1}{R_0}.\nonumber
\end{equation}

Once again, for $R_0 < 1$ the smallest root of equation~\eqref{eq:extinction} is $q=1$ and extinction is guaranteed. For $R_0 > 1$ the probability of extinction decreases monotonically (i.e., an epidemic becomes more likely) both with increasing $R_0$ and $k$. Therefore, as highlighted in~\cite{superspreaders}, higher levels of host heterogeneity (i.e., smaller $k$) lead to a higher probability of epidemic extinction.

\section{Invasion by a novel strain against an endemic background}
\label{sec:strain_invasion}

Now we consider the multi-strain scenario, in which a new invading pathogen strain $i$ is introduced into a population that already hosts a resident endemic strain $r$. We apply a branching process approximation to calculate the probability that strain $i$ establishes itself within the community, i.e., avoids extinction, but note that the approximation used must necessarily account for the impact of the resident strain on the capacity of the invader to infect individuals. We note that Leventhal \emph{et al.} \cite{leventhal2015} take a similar branching process-based approach, but consider network structure explicitly, rather than allowing individual reproductive capacity to vary in a more general way (as is the case here). For simplicity, we assume that the background infection is in some quasi-stationary distribution ({\em sensu} Bartlett \cite{bartlett1957}). We also assume that the population is large, so that given the presence of the resident strain $r$, there remain sufficient susceptible individuals that the introduction of the invading strain $i$ can be approximated by a branching process, and that early in the outbreak of $i$, the fraction of the population that are susceptible does not change substantially.

\begin{figure*}
\begin{center}
  \includegraphics[width=0.75\textwidth]{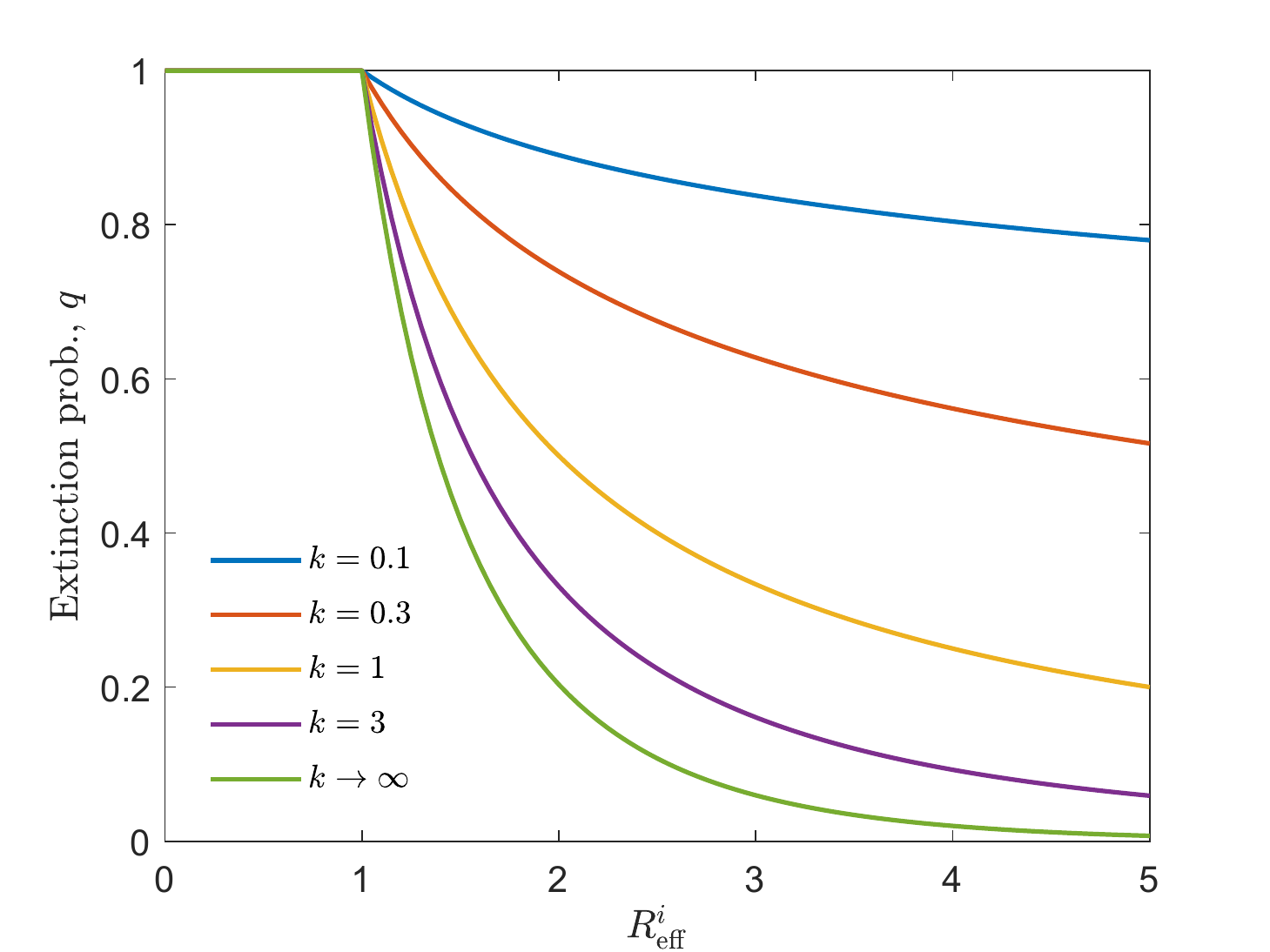}
\caption{Probability of extinction, $q$, for a new invading strain entering an endemically infected population as a function of the effective reproduction number $R_{\mathrm{eff}}^i=R_0^i/R_0^r$ for varying dispersion parameter $k$.}
\label{fig:extinction_probability}       
\end{center}
\end{figure*}

One difference between this scenario and the one considered in the previous section is that, assuming perfect cross-immunity, the presence of a resident strain $r$ diminishes the pool of available susceptibles presented to the invading strain $i$, thus impeding its epidemic potential. Specifically, in our multi-strain scenario, infectious individuals come into contact with (other) individuals according to a Poisson process in the normal way, but rather than those individuals necessarily being susceptible (as in the na{\"i}ve case), they are only susceptible with probability (approximately) $\bar{S}/(N-1)$, where $\bar{S}$ is the 
mean endemic susceptible population, i.e., $\bar{S}=N/R_0^r$, and $N$ is the population size. (For convenience, we note that $N \approx (N - 1)$ when $N$ is large, and thus discard the $-1$ hereafter. Thus, $\bar{S}/N$ is the susceptible fraction of the population.) Consequently, given the presence of the endemic strain, we find that the effective reproduction number, $R_{\mathrm{eff}}^i$, of the invading strain $i$ at the time of its introduction is rescaled accordingly:
\begin{equation}
    R_0^i\rightarrow R_{\mathrm{eff}}^i = R_0^i\,\frac{\bar S}{N}=\frac{R_0^i}{R_0^r}.
\end{equation}
  $R_0^i$ and $R_0^r$ are the mean basic reproduction numbers of the invader and resident strain, respectively, as measured in the absence of the other. Note that this does not rely on a specific formulation around other disease dynamics, e.g., the presence of a recovered class and how waning immunity or replenishment of susceptibles occurs. To calculate the extinction probability of the invading strain in this case we need only replace $R_0$ in equation~\eqref{eq:extinction} with $R_{\mathrm{eff}}^i = R_0^i/R_0^r$, and solve numerically. The results are shown in Figure~\ref{fig:extinction_probability} where we effectively reproduce Figure 2b) of~\cite{superspreaders} with the $x$-axis rescaled from $R_0\rightarrow R_{\mathrm{eff}}^i$. Accordingly we now find that the epidemic threshold condition becomes $R_{\mathrm{eff}}^i > 1$ such that extinction of the invading strain is guaranteed for $R_0^i < R_0^r$. Once again, we observe that the extinction probability decreases monotonically with increasing $R_{\mathrm{eff}}^i$ and, as in~\cite{superspreaders}, that increasing individual heterogeneity (decreasing $k$) favours extinction. 

To verify the utility of this branching process approximation for strain invasion against an endemic background, and test the conditions under which it can be applied with confidence, we performed a simulation study. Full details appear in Appendix A. Exact simulations of a SIRS-type model with two strains and heterogeneity in individual reproductive capacity identical to that considered here were produced across a range of population sizes and parameter choices. The simulation results demonstrated that the branching process approximation was accurate for sufficiently high population sizes (e.g., $N \geq 500$), and that this accuracy was robust to choices of $k$, $R_0^r$, and $R_0^i$.

\section{Probability of strain invasion under control scenarios}
\label{sec:control}

We now consider the impact of control on the invasion (or extinction) probability of a new strain entering an endemic population focussing separately on several key factors: first, the mechanism by which control is implemented and its effect on the distribution of reproductive potentials $\nu$ among members of the infected population; second, the timing of the introduction of control and the implications for the background population dynamics; and, finally, the relative efficacy of intervention measures when applied to the resident strain $r$ and the invading strain $i$, which we anticipate enjoys some level of resistance to control (e.g., drug resistance). 

\subsection{Control policy}

First, we consider two separate types of control measures and their impact (following \cite{superspreaders}): 
\begin{enumerate}
    \item Uniform, population-wide control where each individual in the population experiences a uniform reduction in their transmission potential $\nu$. As a result the reproductive potential $\nu$ of every member of the population is reduced by a factor $1-c$ where $c$ is the level of control: $\nu^{\mathrm{pop}} = (1-c)\nu$.
    \item  Random, polarized control where a fixed proportion $c$ of the infected population is neutralized completely ($\nu^{\mathrm{ind}}= 0$) and the remaining $(1 - c)$ individuals are unaffected ($\nu^{\mathrm{ind}} = \nu$).
\end{enumerate} 
 From the description above we see that for the same level of control $c$, the mean effective reproduction number in the presence of control $R_{\mathrm{eff},c}^i = (1-c)R_{\mathrm{eff}}^i$ is the same in both cases. Importantly however, in general the polarized control policy acts to increase the spread (i.e. heterogeneity) in individual infectiousness, $\nu$, among the infectious cohort, whilst uniform control reduces it.
 
 We emphasise that control here is something that impacts infectious individuals rather than susceptible individuals; specifically, under polarized control, the controlled individual becomes infected, but then produces zero offspring. This definition follows from our choice to model the outbreak as a branching process where we only track the infectious cohort.

For the uniform and polarized control scenarios described above, the probabilities of epidemic extinction $q^{\mathrm{u}}$ and $q^{\mathrm{p}}$ are respectively determined according to the following equations:
\begin{align}
    \mbox{\underline{Uniform control}:} \qquad q^{\mathrm{u}} &= \left(1 + (1-c)\frac{R_{\mathrm{eff}}^i}{k}(1 - q^{\mathrm{u}})\right)^{-k};\nonumber\\
    \mbox{\underline{Polarized control}:} \qquad q^{\mathrm{p}} &= c + (1 - c)\left(1 + \frac{R_{\mathrm{eff}}^i}{k}(1 - q^\mathrm{p})\right)^{-k}.\label{eq:extinction_control}
\end{align}
In the uniform control case all individuals are impacted equally and the extinction equation resembles equation~\eqref{eq:extinction}, only now the effective reproduction number $R_{\mathrm{eff}}^i$ has been appropriately scaled according to the level of control. In the polarized case, a fixed proportion of infected individuals $c$ are guaranteed to produce no subsequent infections (i.e., those branches of the infection process become extinct) whilst the extinction probability of the remaining $(1-c)$ infected individuals remains unaltered. Since polarized control acts to increase population heterogeneity and uniform control reduces it, we expect that $q^{\mathrm{p}} > q^{\mathrm{u}}$ for all $c\in (0,1 - 1/R_{\mathrm{eff}}^i)$, which, in fact, can be proven analytically~\cite{superspreaders}.

For now, we assume that control measures are  equally effective against both the resident strain $r$ and the invading strain $i$ and that they are implemented coincident with the emergence of the invading strain. 
We relax these assumptions in the upcoming sections.

\begin{figure*}
\begin{center}
  \includegraphics[width=0.75\textwidth]{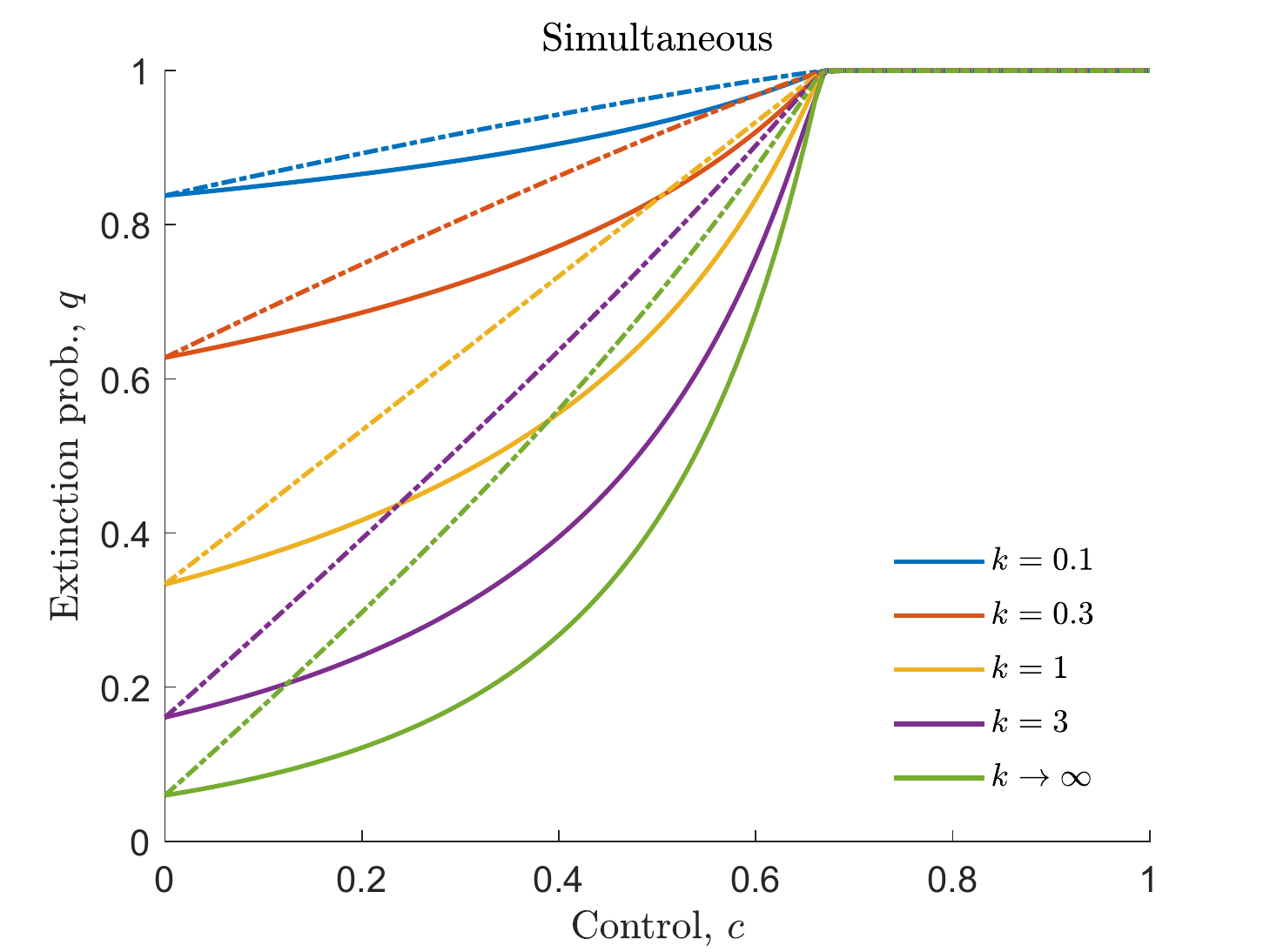}
\caption{Comparison of the probability of extinction, $q$, for a new invading strain entering an endemically infected population under a uniform (solid) and polarized (dot-dashed) control policy as a function of the level of control $c$ and varying dispersion parameter $k$. Here, we have set $R_0^i = 5$ and $R_0^r = 5/3$ which gives a critical control level $c_{\mathrm{crit}} = 1 - R_0^r/R_0^i = 2/3$. In this figure we have assumed that control has been implemented coincident with the emergence of the invading strain $i$ and that it is equally effective against both strains.}
\label{fig:simultaneous_control}       
\end{center}
\end{figure*}

In Figure~\ref{fig:simultaneous_control} we compare the extinction probabilities of the invading strain $i$ under the uniform (solid) and polarized (dot-dashed) control policies discussed above where, for reference, we have set $R_0^i = 5$ and $R_0^r = 5/3$ such that $R_{\mathrm{eff}}^i = 3$. Similar to before, in the uniform control case the effect of adding control is to rescale the extinction curves presented in Figure~\ref{fig:extinction_probability}, i.e. $R_{\mathrm{eff}}^i\rightarrow (1-c)R_{\mathrm{eff}}^i$. For both control policies, we see that extinction of the invading strain is guaranteed for $c \geq c_{\mathrm{crit}} = 1 - 1/R_{\mathrm{eff}}^i$ and $q^{\mathrm{p}} > q^{\mathrm{u}}$ for all $c \in (0, 1 - 1/R_{\mathrm{eff}}^i)$. That is, the polarized control policy, for which the extinction probability $q$ increases almost linearly with $c$ for $c < c_{\mathrm{crit}}$, outperforms the uniform control policy, in terms of preventing the establishment of the invading strain.

Note that applying control to the resident strain will cause the susceptible background to shift. How quickly this occurs depends upon how control is applied: if it were applied to all existing infected individuals immediately, the background would likely shift quite quickly, causing the branching process approximation for invader success to be inaccurate. If control is instead applied only to new infectious individuals, the susceptible background will shift more slowly, and the branching process approximation may remain relatively accurate. However, as control coincident with the introduction of the invading strain is somewhat unrealistic, we will not investigate this in more detail. Instead, in the next section we consider the more realistic scenario, where control precedes the introduction of the invader.

\subsection{Control timing}


In the previous section we assumed that control measures were implemented coincident with the emergence of the invading strain $i$. There it was understood that any intervention measures applied to the resident strain $r$ will not have had sufficient time to suppress the prevalence of the endemic strain $r$ and replenish the susceptible pool, $S$. We now consider an alternate (more realistic) scenario in which control has been implemented over the long-term to the endemic disease population prior to strain $i$'s emergence, i.e., at $t < 0$. In this case we assume that control has been applied sufficiently in advance that the population dynamics have had sufficient time to re-equilibrate.

The major difference between the two scenarios described above is the number of susceptibles greeting the invading strain $i$: controlling the resident strain $r$ over the long term replenishes the susceptible pool, $S$ and, if the level of control is sufficient, eventually the resident strain is eliminated and the whole population becomes susceptible, i.e., $S(t=0)\rightarrow N$. Therefore, in the prior control scenario the invading strain $i$ has an increased effective reproduction number which saturates upon elimination of the resident strain:
\begin{equation}
    R_{\mathrm{eff}}^i = \frac{R_0^i}{\mathrm{max}[(1-c)R_0^r, 1]}.\label{eq:Reff_prior}
\end{equation}
Note that this expression does not yet take into account the effect of control applied to the invading strain $i$ itself; for that we need to substitute~\eqref{eq:Reff_prior} into 
the extinction probability equation~\eqref{eq:extinction_control}. In doing so, we observe that for the uniform control case the factor $(1 - c)$ cancels and the extinction probability $q^{\mathrm{u}}$ is independent of the level of control $c$, at least until the resident strain becomes extinct at $c > 1 - 1/R_0^r$. Below this threshold ($c < 1 - 1/R_0^r$), the increase in the susceptible population awaiting strain $i$ (as a result of controlling the resident strain $r$) negates the effects of any intervention measures that are applied to the invading strain. Therefore, in this regime, uniform control is ineffective at preventing the invasion of the novel strain (though we should certainly keep in mind that such measures would still reduce the overall/combined disease prevalence, relative to a scenario with no control).
A similar cancellation does not occur for the polarized control case and given that $q^{\mathrm{p}} > q^{\mathrm{u}}$ we anticipate that control will always be effective in this case.

\begin{figure*}
\begin{center}
  \includegraphics[width=0.75\textwidth]{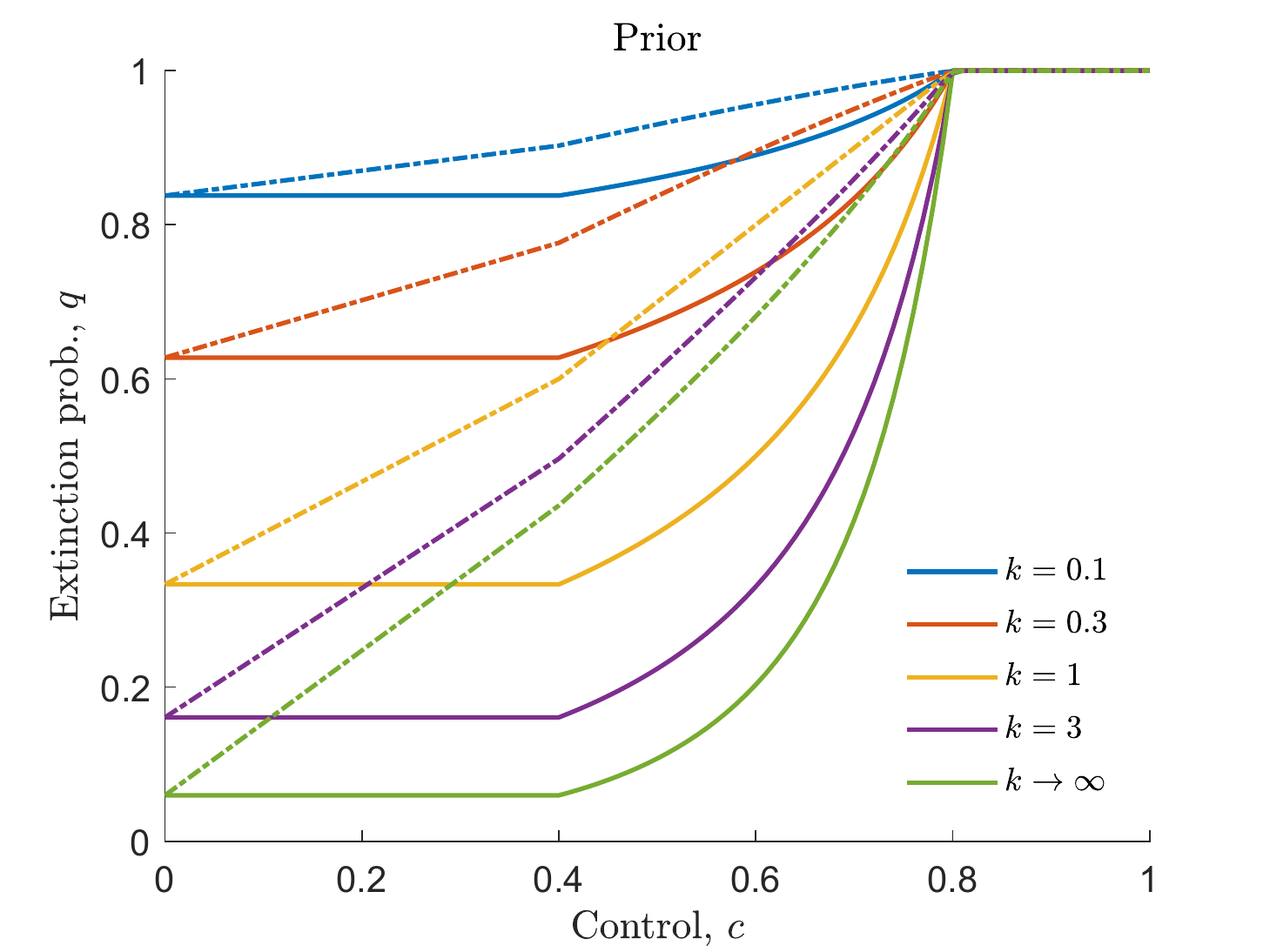}
\caption{
Same as Figure~\ref{fig:simultaneous_control} for the prior control policy.
}
\label{fig:prior_control}       
\end{center}
\end{figure*}

A plot of the extinction probability is given in Figure~\ref{fig:prior_control} where again we compare the extinction probabilities of the invading strain $i$ under the uniform (solid) and polarized (dot-dashed) control policies discussed above. For ease of comparison with Figure~\ref{fig:simultaneous_control}, we have once again set $R_0^i = 5$ and $R_0^r = 5/3$. Immediately we observe that for $c < 1-1/R_0^r$ uniform control measures have no impact on the invasion prospects of strain $i$. Not until we enter the second regime, $1-1/R_0^r \leq c < 1 - 1/R_0^i$, for which the resident strain $r$ has been eradicated prior to the emergence of the invading strain $i$, does increasing the level of control $c$ increase the extinction probability of strain $i$. Conversely, polarized control is always effective for $c > 0$, and becomes increasingly so for $c > 1 - 1/R_0^r$.

\subsection{Differential control}

In the preceding analysis we assumed that the control measure being implemented is equally effective against both the resident strain $r$ and the invading strain $i$.
A more general scenario to consider is when the control measure that is implemented differentially impacts the two strains. Indeed, if the invading strain is a mutant variant of the wild-type resident strain (as expected in the AMR case) it is reasonable to expect that this new strain would enjoy some level of resistance to active control measures. Therefore, in this section we generalize the approach taken above by assuming that for a fixed level of control $c$ applied to the background resident or endemic strain $r$, only a fractional component $\alpha c$ --- where $0 \leq \alpha \leq 1$ --- inhibits the progress of the invading (possibly drug-resistant) strain.

In this case the extinction equations~\eqref{eq:extinction_control} given above generalize to
\begin{align}
    \mbox{\underline{Uniform control}:} \qquad q^{\mathrm{u}} &= \left(1 + (1-\alpha c)\frac{R_{\mathrm{eff}}^i}{k}(1 - q^{\mathrm{u}})\right)^{-k};\nonumber\\
    \mbox{\underline{Polarized control}:} \qquad q^{\mathrm{p}} &= \alpha c + (1 - \alpha c)\left(1 + \frac{R_{\mathrm{eff}}^i}{k}(1 - q^\mathrm{p})\right)^{-k},\label{eq:extinction_differential}
\end{align}
where we previously observed that the effective reproduction number in the simultaneous and prior control scenarios are given respectively by
\begin{align}
    \mbox{\underline{Simultaneous control}:} \qquad R_{\mathrm{eff}}^i &= \frac{R_0^i}{R_0^r};\nonumber\\
    \mbox{\underline{Prior control}:} \qquad R_{\mathrm{eff}}^i &= \frac{R_0^i}{\mathrm{max}[(1-c)R_0^r, 1]}.\label{eq:R0_timing}
\end{align}
For the simultaneous control case (with $R_{\mathrm{eff}}^i= R_0^i/R_0^r$) the effect of differential control is to simply rescale the control factor $c \rightarrow \alpha c$. TAs a result, in this particular case, the extinction probabilities (for the invading strain) calculated using equations~\eqref{eq:extinction_differential} and~\eqref{eq:R0_timing} will be qualitatively the same as that presented in Figure~\ref{fig:simultaneous_control}, with the $x$-axis $c$ reduced to $\alpha c$. Therefore, in this section we concentrate only on the prior control scenario.

\begin{figure*}
\begin{center}
  \includegraphics[width=0.75\textwidth]{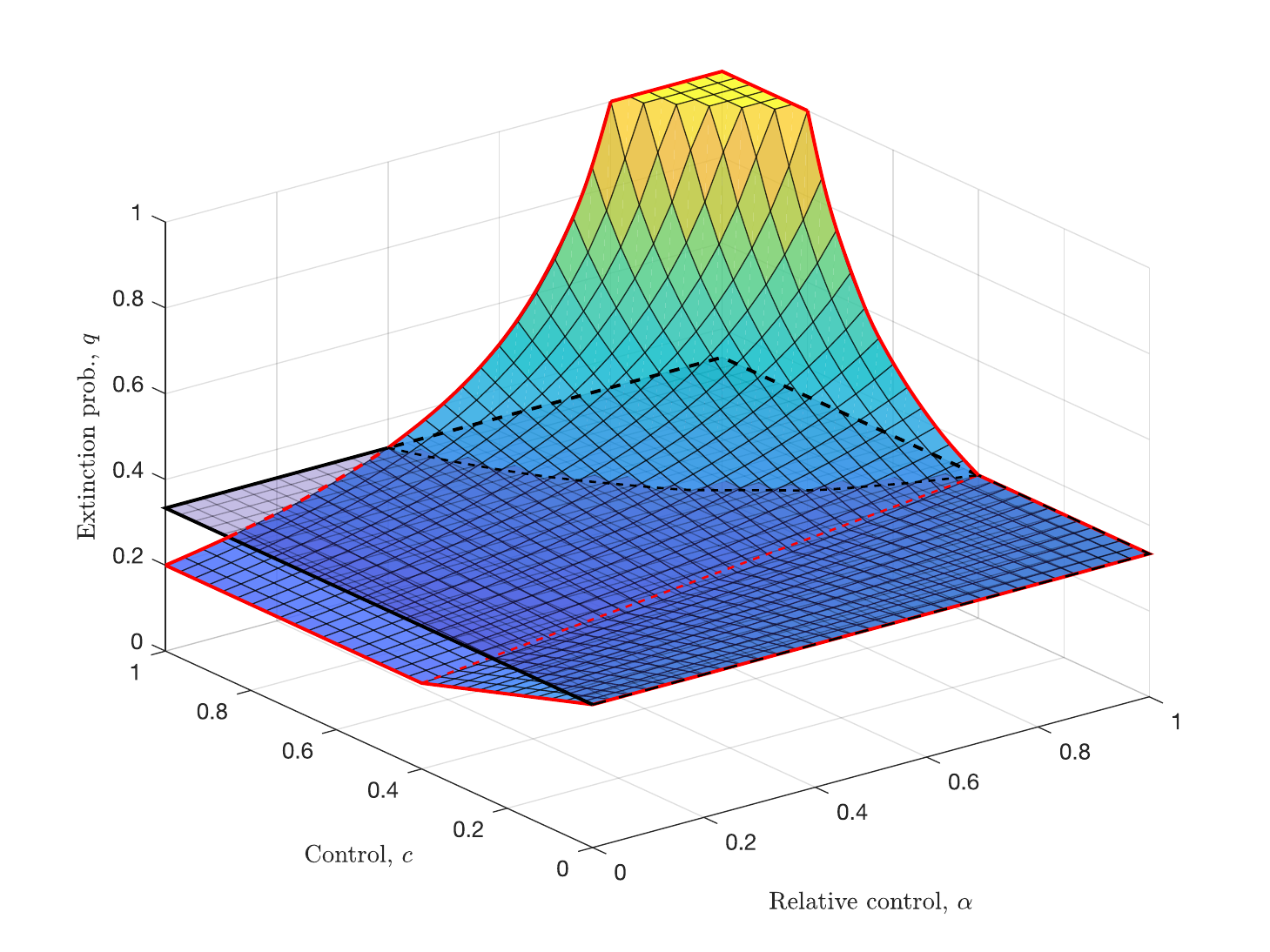}
  \includegraphics[width=0.75\textwidth]{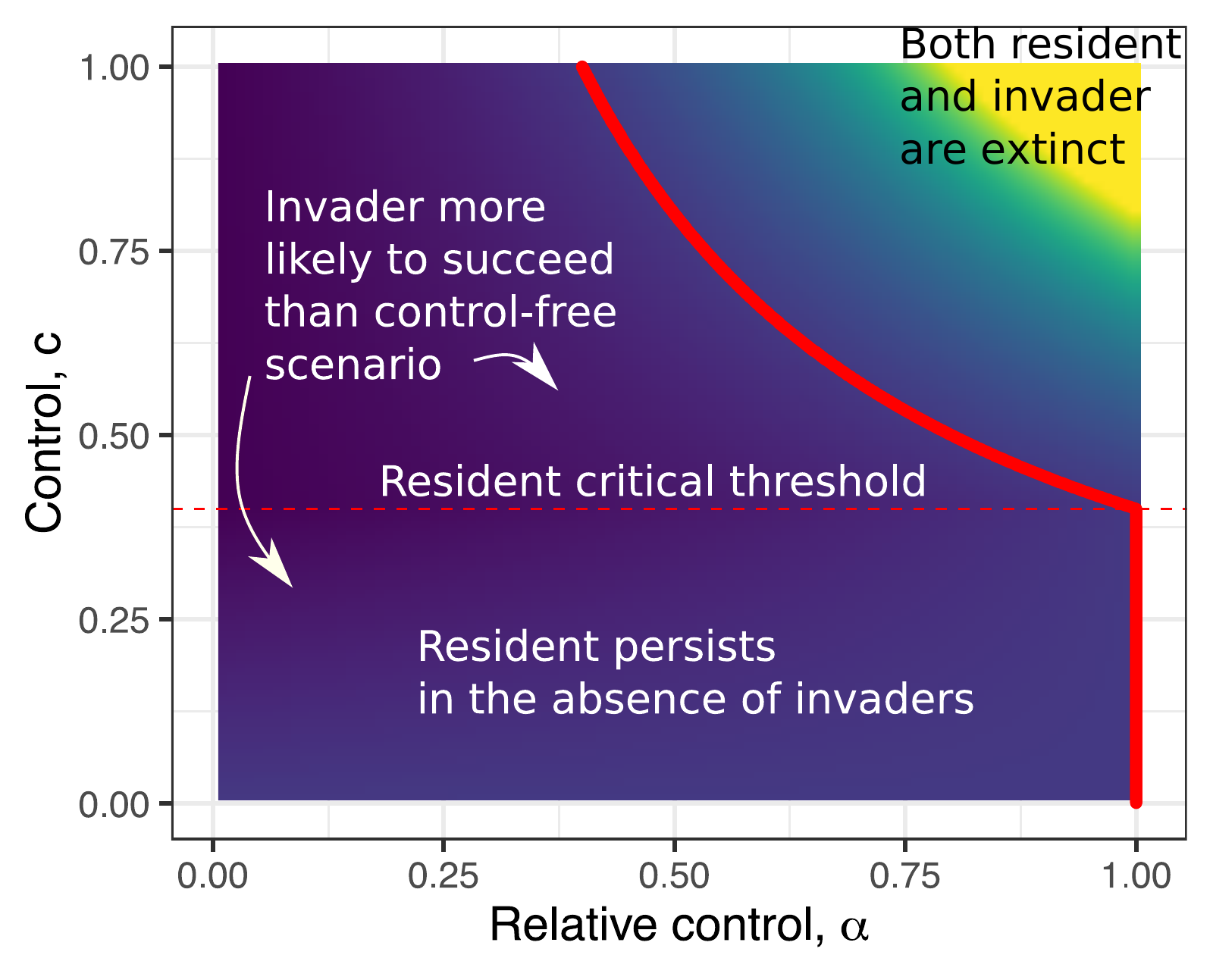}
\caption{Probability of extinction for the invading strain, $q$, under the differential uniform control scenario. (a) Extinction probability varies as a function of the resident control fraction $c$ and the relative control fraction $\alpha$. Here we have taken $k = 1$, $R_0^i = 5$ and $R_0^r = 5/3$, and assumed that control has been implemented prior to the emergence of the invading strain. For comparison, we have also drawn the reference plane for the extinction probability in the absence of control, i.e., for $c = 0$. (b) A schematic describing the meaning of the different regimes that occur. The dashed line represents the critical threshold of the resident strain, i.e., for control below this threshold, the resident may persist in the absence of the invader, for control above the threshold, the resident becomes extinct. The solid red line indicates the boundary of the region where control facilitates invader success relative to the control-free scenario.}
\label{fig:uniform_differential}       
\end{center}
\end{figure*}

\begin{figure*}
\begin{center}
  \includegraphics[width=0.75\textwidth]{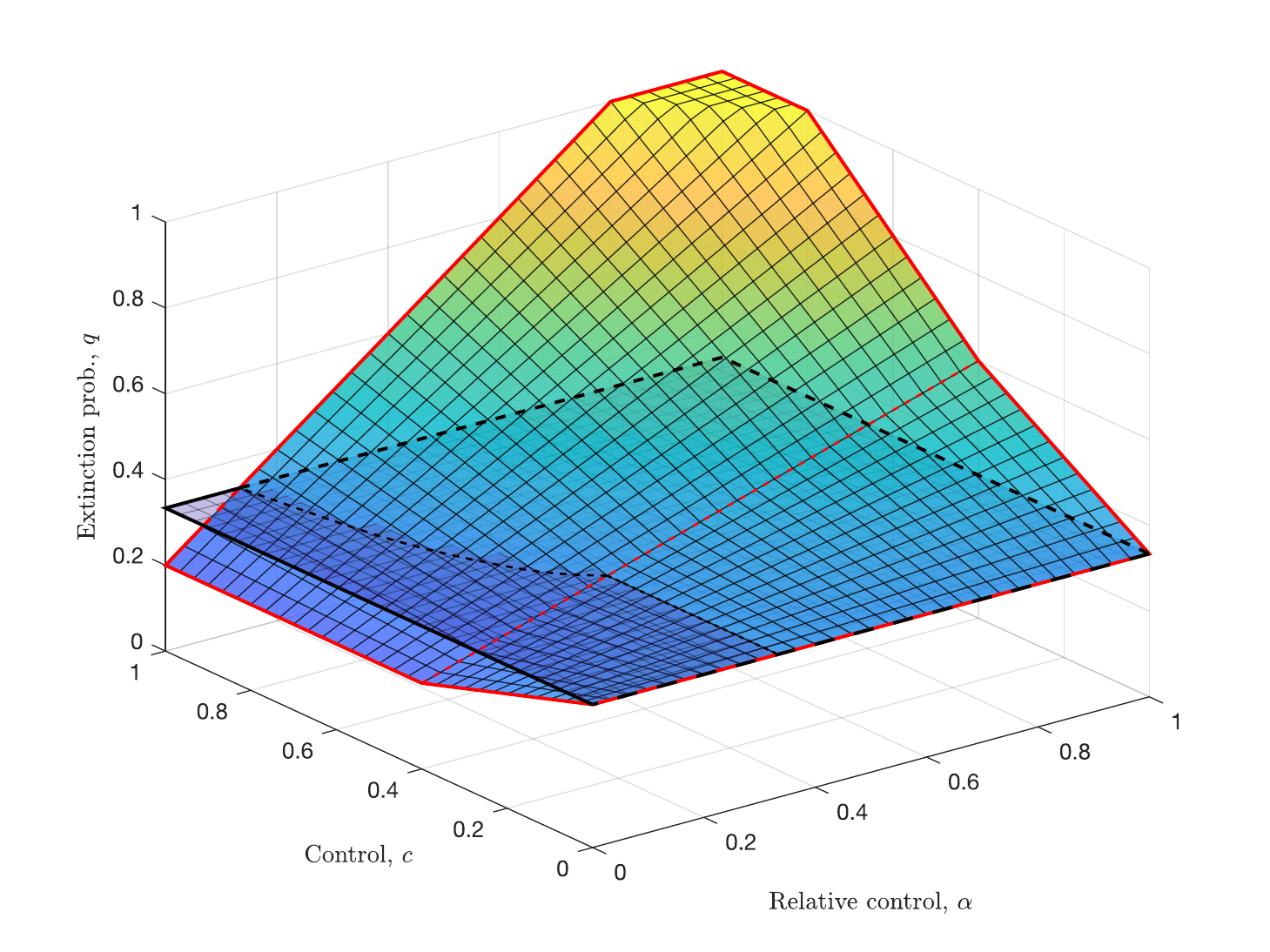}
  \includegraphics[width=0.75\textwidth]{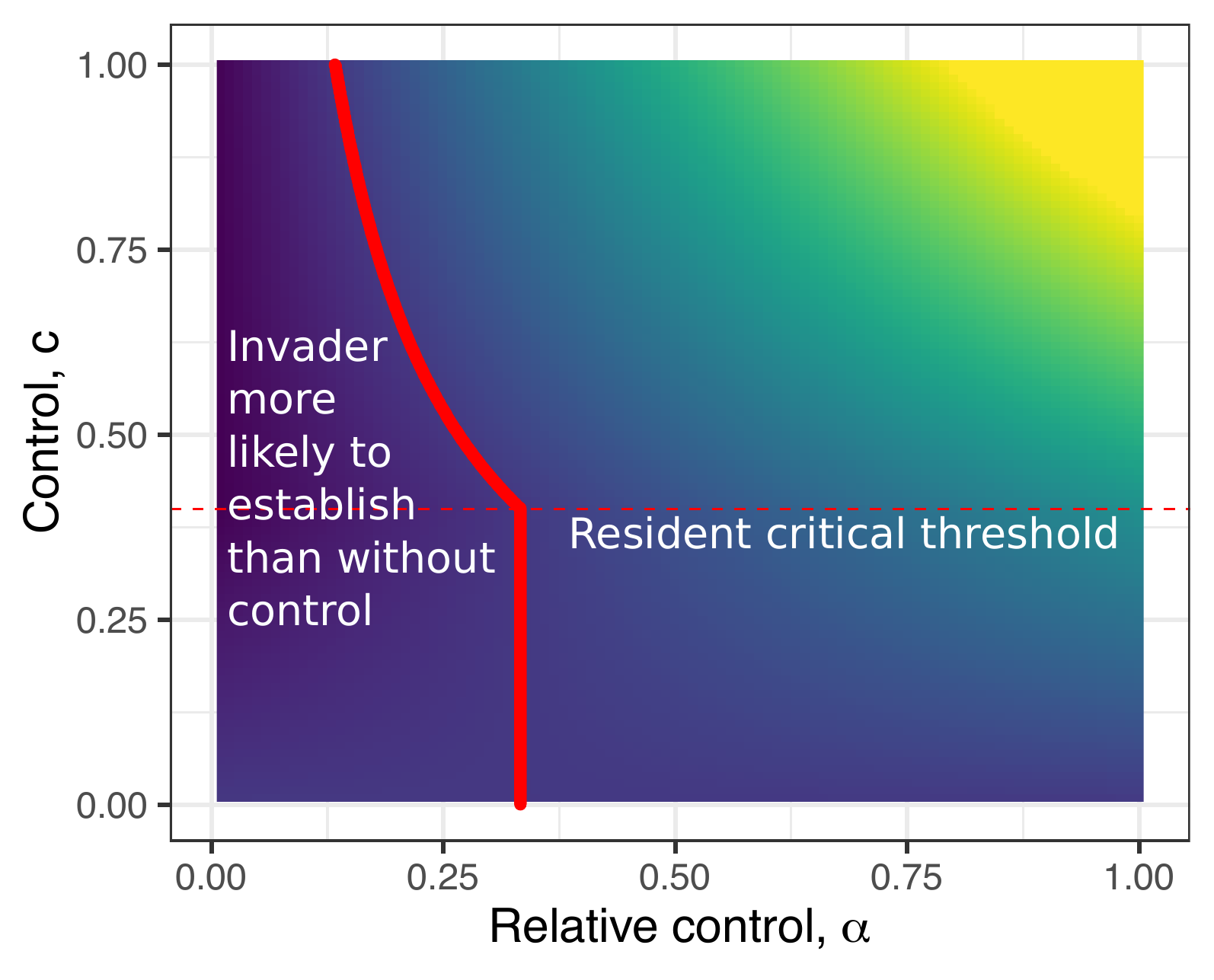}
\caption{
Same as Figure~\ref{fig:uniform_differential} for polarized control.
}
\label{fig:polarized_differential}       
\end{center}
\end{figure*}

If we substitute the expression for the effective reproduction number under the prior control scenario into~\eqref{eq:extinction_differential} we obtain results that are both quantitatively and qualitatively distinct from those found previously (see Figures~\ref{fig:uniform_differential} and~\ref{fig:polarized_differential}). Of particular interest is the combination of uniform control implemented prior to the emergence of the invading strain in the sub-elimination regime $c \leq 1 - 1/R_0^r$ (i.e., below the critical level required to eliminate strain $r$). In this particular instance, the effective reproduction number in the presence of control becomes
\begin{equation}
    R_{\mathrm{eff},c}^i = \frac{(1 - \alpha c)R_0^i}{(1 - c)R_0^r} > \frac{R_0^i}{R_0^r} \quad \mbox{for} \quad 0 < \alpha, c < 1 \nonumber
\end{equation}
such that increasing $c$ will actually increase the effective reproduction number of strain $i$ and enhance its invasion prospects relative to the no control $(c = 0)$ scenario. Similarly, although a closed analytical expression cannot be found, it is also possible to show that an analogous region of parameter space exists in which the application of polarized control prior to the emergence of the invading strain leads to an enhanced probability of invasion, relative to the no control scenario. 

An important corollary to these results is that under the differential control scenario, invasion can occur even when the invading strain is less transmissible than the resident strain, i.e., when $R_0^i < R_0^r$ it is still possible to obtain $R_{\mathrm{eff},c}^i > 1$. 
That is, while implementing control can have substantial positive impacts in terms of reducing the prevalence of a resident disease strain, it may facilitate the emergence of new, drug-resistant strains (that would not have emerged in the absence of control).

To illustrate this finding we solve equation~\eqref{eq:extinction_differential} using the prior control effective reproduction number given in~\eqref{eq:R0_timing} (see Figures~\ref{fig:uniform_differential} and~\ref{fig:polarized_differential}). As a specific example we have taken $k = 1$ (corresponding to the Exponentially-distributed infectious potential $\nu$), but, we point out that the behaviour is qualitatively the same for different values of $k$ --- the difference being that the region for which control promotes invasion by the invading strain increases as $k$ increases.

We observe that the region for which control enhances the invasion prospects of strain $i$ is larger for the uniform control case compared to the polarized control scenario (Figures~\ref{fig:uniform_differential} and~\ref{fig:polarized_differential}). This observation follows from our earlier remark that $q^{\mathrm{p}} > q^{\mathrm{u}}$ meaning that invasion is always less successful in the polarized case. However, even in the polarized control case we still observe enhanced invasion probabilities provided the level of relative control $\alpha$ is sufficiently small, which in the event of multi- or even extensively-drug-resistant pathogens could well be the case.

Finally, we reiterate that while control may in some cases facilitate the establishment of an invader (that may not have established in a control-free environment), the application of control always necessarily reduces overall disease prevalence. Moreover, note that, except in the case where the endemic disease has been eradicated due to control efforts, introduction in competition with an endemic background is always more challenging than introduction into a na{\"i}ve population. Even in the extreme case in which the level of differential control $\alpha\rightarrow 0^+$, the effective reproduction number $R_{\mathrm{eff},c}^i < R_0^i$. Hence the presence of an endemic strain always acts to protect the host population from strain invasion provided there is some level of cross-immunity.

\section{Conclusion}
\label{sec:conclusion}

The number of phenotypically distinct strains exhibiting varying levels of resistance to antimicrobial treatments continues to rise, with new strains, enjoying even more extensive levels of resistance continually emerging. 
Due to the stochastic nature of disease transmission, the successful emergence of a new strain is not certain, even when that strain is more fit than any pre-existing strains circulating in the population. Understanding the relationships between relative strain transmissbility, population heterogeneity, and the impact of different control measures on strain emergence will enable more effective modelling of multi-strain disease dynamics in the face of challenges such as AMR.

Our manuscript presents three key advances. First, we extend previous work on disease emergence in an infection-na{\"i}ve population \cite{superspreaders} to address emergence of novel mutant strains competing against resident endemic strains. This is a simple but powerful theory that can inform multi-strain modelling projects going forward (e.g., \cite{meehan2018,mcbryde2017}), including deterministic approximations, which can be applied once the outbreak has successfully established \cite{rebuli2017}.
Second, we demonstrate that the time at which control measures are implemented has a substantial impact on the effectiveness of control for preventing the emergence of new strains, and that this varied by the chosen means of control.
Third, in the case of differential control, where the invading strain shows some level of resistance to the applied control measure, we found regions of parameter space for which control was detrimental and increasing $c$ increased the invasion probability of the new strain. Hence, when control impacts the invading strain differently to the resident, this can (in some cases) facilitate the emergence of the invading strain, relative to the control-free case.

 We observed that, for equal levels of effective coverage, polarized control measures that completely neutralize a random fraction of the population (e.g., perfect quarantine of a proportion of infected individuals) consistently outperform imperfect control measures that partially control the entire population (e.g., vaccination that reduces, but does not eliminate, the susceptibility of all individuals). That is, random, polarized control provides greater protection against potential strain invasions. This is again consistent with results around invasion in na{\"i}ve populations \cite{superspreaders}. This could provide an important consideration for managers choosing control strategies when concerns exist around AMR. Of course, more targeted control measures directed towards the most highly infectious members of the host population --- so-called `super-spreaders' --- would outperform both the random and uniform control scenarios considered here; however, identifying such groups \textit{a priori}, if possible, may be difficult in practice. 

For control timing, we considered two separate scenarios: implementing the various control strategies (i.e., random polarized and uniform imperfect) only upon emergence of the invading strain; and doing so prior to its emergence, such that the population dynamics have had sufficient time to re-equilibrate. We note that the latter case, where control exists prior to the introduction of the invading strain, is likely to be more realistic, given the desire to control existing outbreaks. The difference between these two cases is the number of susceptible individuals made available to the newly introduced invading strain. In the simultaneous case, the susceptible pool remains depleted by the presence of the resident strain such that the invading strain is inhibited by both a reduced $S(t=0) = N/R_0^r$ and the added burden of control. In the prior scenario, the susceptible population has been replenished by control measures applied to the resident strain, $S(t=0) = N/((1-c)R_0^r)$. In this latter case, the effects of uniform control on the invading strain are negated by the increase in available susceptibles: $R_{\mathrm{eff},c}^i = R_{\mathrm{eff}}$. Only once the level of control is sufficient to eradicate the resident strain, i.e., $c > 1 - 1/R_0^r$ (and the susceptible population then becomes the whole population, $N$) does uniform control start to impede the invasion prospects of the invading strain. Conversely, implementing a random, polarized control strategy is always beneficial provided the control measure is equally effective against both the resident and invading strains. 


One limitation of our analysis is that we assume that the susceptible individuals are the same, and mix homogeneously, so that they are equally likely to become infected. If this were not the case, the resident infection would likely impact which susceptibles are available to the invader, requiring a more complex modelling framework. For example, given that both infection potential and susceptibility are functions of the number of contacts an individual experiences, these quantities may be naturally correlated. Leventhal \emph{et al.}~\cite{leventhal2015} approach this problem by imposing an explicit contact-network structure on the study population, such that heterogeneity in susceptibility was directly linked to the heterogeneity in infectiousness through the (fixed) number of contacts shared by each individual. In this context, invasion events were found to be less likely because potential `super-spreading' hosts (or hubs in the network) were often already infected with the resident strain meaning they were unable to perpetuate the spread of the invading strain, whilst simultaneously providing protection to susceptible satellite nodes. Given these findings, it would be interesting to consider more general correlation structures between susceptibility and infectiousness and their effect on strain invasion potential in a generalized branching process framework.


Throughout our analysis, we considered the potential for heterogeneous individual reproductive capacity (i.e., the capacity for `super-spreaders') among members of the host population. While this does not influence the impacts of control qualitatively, we emphasize that under the branching process approximation it does impact how likely the invading strain is to establish. In particular, invaders are less likely to establish when individual reproductive capacities are more highly dispersed. However, we note that this heterogeneity also has a subtler effect: the error between simulated epidemic results and the branching process approximation varies in both magnitude, and direction, with $k$ (Figure \ref{fig:grid_errors}). One clear possibility is that if a `super-spreader' individual emerges early in an outbreak, it may allow a strain to establish itself when it otherwise would not have. In addition, since the heterogeneity also influences the resident strain, increased heterogeneity both enables the natural extinction of the resident strain (Figure \ref{fig:extTimes}), and more generally variation of the initial susceptible proportion faced by the invader away from $N/R_0^r$.

Applying a branching process approximation for the emergence of a new strain in the presence of an endemic background, rather than into a na{\"i}ve population, requires care. Our simulation study determined that the approximation was quite robust even for relatively small populations (e.g., $N \geq$ 500). However, there was some variation due to different parameter choices at smaller population sizes. In particular, accuracy varied when heterogeneity in individual reproductive capacity was high, and when parameter choices caused the endemic disease to be at risk of natural extinction (even without the influence of the invader). We also emphasize that the branching process approximation is limited in that it can only quantify the probability of initial strain extinction: there is likely some additional probability of subsequent fade-out due to depletion of susceptibles (e.g., {\em sensu} Ballard \emph{et al.} \cite{ballard2016}). A separate analysis would be necessary to investigate this phenomenon.

One key assumption we make here is that, in effect, we require that (functional) mutations occur sufficiently rarely that the new strain will either establish, or become extinct, before a repeat mutation can occur. 
If instead there was a possible ancestral relationship between the resident strain and the invader, where the resident effectively fuels the invading cohort with continual reinforcements, invasion dynamics would likely be different. It would be interesting then to re-evaluate our results with a coupling between the two strains: our expectation is that with a perpetual supply of reinforcements, replacement with any fitter (i.e., $R_0^i > R_0^r$) variants is guaranteed --- it just a question of when. Further, assuming an ancestral relationship between the competing strains we could also model multiple epochs (i.e., we would not terminate the simulation after a single replacement event has occurred) and follow the evolution of the pathogen. These considerations are left as avenues for future research.

The results of this analysis point to the role of endemic infection as an immunizing agent and its ability to impede the emergence of new exotic strains. 
In addition to population-wide effects, this has important implications at the individual level, where the use of broad-spectrum antibiotics can inadvertently eliminate colonizing bacterial infections (e.g. \textit{Staphylococcus aureus}), and thus further highlights the importance of antibiotic stewardship.
Perhaps the most critical outcome of this work is the potential for the invading strain to become established even when it is less transmissible than the resident due to differential control. The emergence of AMR strains in the presence of endemic backgrounds is of substantial concern in many disease settings (particularly e.g., Tuberculosis), and the impact of differential control highlighted here should be considered in any future control efforts as well as efforts to model multi-strain dynamics in these systems.

\section{Acknowledgements}

This research was supported by a seed funding grant (to M.T.M. and R.C.C.) from the NHMRC Centre of Research Excellence in Policy Relevant Infectious disease Simulation and Mathematical Modelling ($\textrm{PRISM}^2$). R.C.C. received funding from the Data to Decisions Cooperative Research Centre (D2DCRC). This work was supported with supercomputing resources provided by the Phoenix HPC service at the University of Adelaide.

\newpage

\section{Appendix A: Simulation study}

To determine conditions under which the branching process approximation for invader establishment success could be confidently applied, we performed a simulation study. 

\begin{figure*}
\begin{center}
  \includegraphics[width=0.75\textwidth]{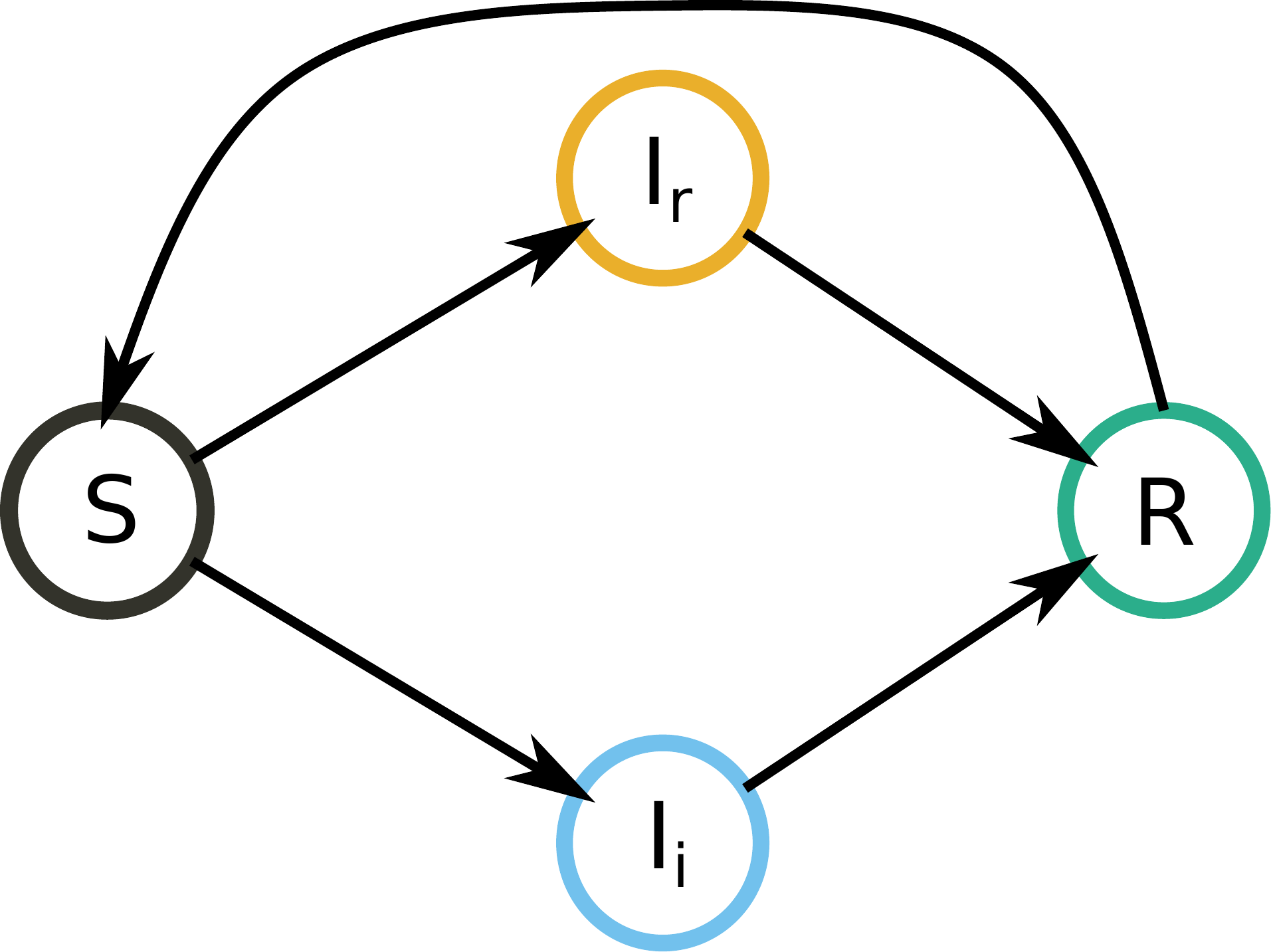}\\
  \includegraphics[width=0.75\textwidth]{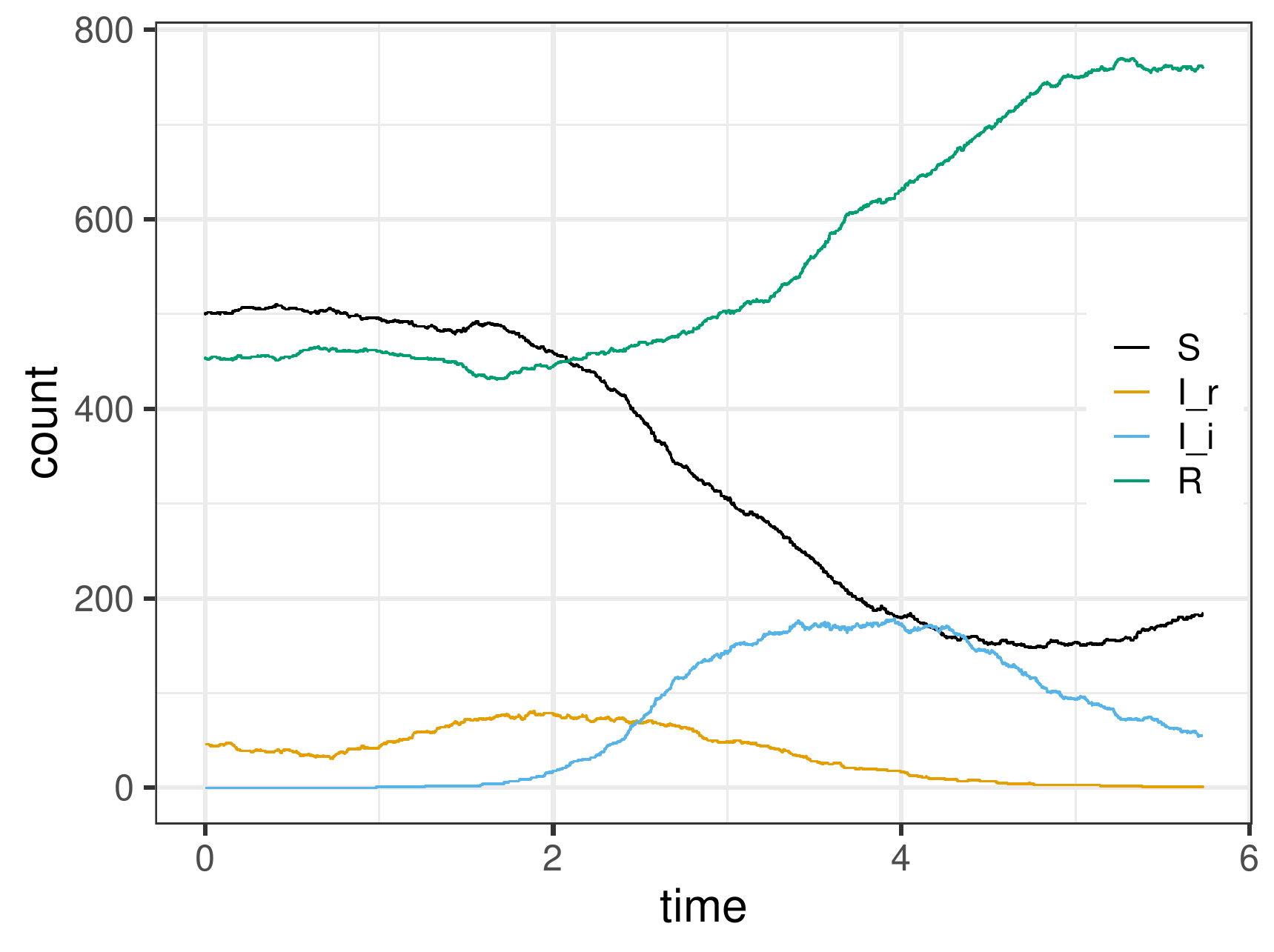}
\caption{(Top) Model schematic for simulated example model. Susceptible individuals become infectious according to a Poisson process; infectious individuals remain infected for exactly one time unit, deterministically. Recovered individuals have their immunity wane and become susceptible according to a Poisson process. (Bottom) Example simulation of this process, with $N = 1000, k = 1, \gamma = 1, R_0^r = 2, R_0^i = 4$. The simulation is initialised in an approximate equilibrium state for the resident disease, and a single infectious individual of the invader strain is introduced at $t = 1$. The simulation ends when the resident strain is extinct.}
\label{fig:eg_model}       
\end{center}
\end{figure*}

We considered a relatively simple stochastic epidemic model (Figure~\ref{fig:eg_model}), in which each of the $N$ individuals in the population are either: susceptible ($S$); infectious with the resident strain ($I_r$), or the invader ($I_i$); or recovered ($R$). Transitions in this model are somewhat unconventional. We assume the infectious duration of each individual is \emph{constant}, rather than random, with value $1/\gamma$ (and for simplicity we set $\gamma = 1$, without loss of generality). Each infected individual $j$ (when infected) has their own individual reproductive capacity, $\nu_j$, Gamma distributed with mean $R_0^s$ and dispersion parameter $k$, where $s$ is the strain they are infected with. Infection events occur according to a Poisson process, with rate $\frac{S }{N-1} \sum_{j \in \textrm{infected}} \nu_j$, i.e., where one would normally expect a $\beta I$ transmission term when infection rates are homogeneous, we instead have $\sum_{j \in \textrm{infected}} \nu_j$. The combination of the deterministic infectious period and the Gamma-distributed individual reproductive capacity results in this situation being equivalent to that considered throughout the manuscript. Recovered individuals have waning immunity, occurring according to a Poisson process with rate $\eta R$.

While this process is not Markov, it can nonetheless be simulated through a small modification of the standard Doob-Gillespie stochastic simulation algorithm (Algorithm 1). The key step is to record the recovery times $r_j$ and individual reproductive capacities ($\nu_j$) of each infected individual in a first-in-first-out (FIFO) queue data structure. Then, rather than generating the time of the next event as in a standard Markov process, we generate a candidate time for the next infection or recovery event, and check if it is due to occur before the next recovery that is due (i.e., the first $r_j$ in the queue). If the candidate infection time is before the next recovery, infection occurs, otherwise the recovery occurs. 


For consistency with the situation described in the main article, we initialise the simulations at $t=0$ in an approximate equilibrium state for $I_r$ (i.e., such that the average transition rates into and out of each compartment are approximately equal; effectively a moment-closure approximation for the mean of the process), and introduce a single $I_i$ at $t=1$. Full details appear in Algorithm 1. An example realisation of this process appears in Figure \ref{fig:eg_model}.

\newpage


Simulations were produced for:
\begin{itemize}
    \item $N$ ranging between 100 and 1000.
    \item $k$ taking values 0.1, 0.3, 1, 3, and 1000 (as in the main manuscript, but with 1000 in place of $\infty$).
    \item $R_0^r$ taking values 2, 3, 4 and 10.
    \item $R_{\textrm{eff}}^i$ taking values 1, 1.1, 1.5, 2, 3, 4 and 5.
\end{itemize}
At least 10,000 simulations were produced for each combination of parameters. $\eta = 1/10$ was used throughout. Simulations were terminated at the later of time $t = 5$, and the time when one or other of the strains became extinct. The $t = 5$ minimum was established as in some cases (particularly with small $N$) the resident strain could become extinct with little influence from the invading strain, with the invading strain becoming extinct shortly after -- in which case the invading strain should not be recorded as successfully becoming established.

Note that we do not consider extinction of the invader at a later time, e.g., epidemic fade out in the first trough (as in \cite{ballard2016}), unless that fade out occurs within the $t \leq 5$ interval.

\begin{algorithm}
\tiny
\KwData{N, k, $R_0^r$, $R_0^i$, $\eta$, $\gamma = 1$, $T_{\textrm{max}}$}
\KwResult{Produces an exact simulation of the epidemic process.}

initialise:
$t = 0$\;
$S(0) = \texttt{floor}(\frac{1}{R_0^r}  N)
$\;
$R(0) = \texttt{floor}(
  \frac{(1 - \frac{1}{R_0^r})}{(1 + \frac{\eta}{\gamma})}
      N
            )$\;
$I_r(0) = N - S(0) - R(0)$\;
$I_i(0) = 0$\;

$\mathscr{U}^r = \texttt{queue}()$;
$\mathscr{V}^r = \texttt{queue}()$;
$\mathscr{U}^i = \texttt{queue}()$;
$\mathscr{V}^i = \texttt{queue}()$\;
\For{$j \in 1,\dots,I_r(0)$}{
$r^r_j \sim \textrm{Uniform}(0,1)$\;
$\texttt{push}(\mathscr{U}^r, r^r_j)$\;
$\nu^r_j \sim \textrm{Gamma}(\gamma R_0^r/k,k)$\;
$\texttt{push}(\mathscr{V}^r, \nu^r_j)$\;
}
$\mathscr{U}^r \leftarrow \textrm{sort}(\mathscr{U}^r)$\tcp*{Existing infectious individuals at time 0 must recover between time 0 and 1, uniformly.} 
$\texttt{push}(\mathscr{U}^i ,\inf)$\;
$\texttt{push}(\mathscr{V}^i ,0)$\;

\While{$t < T_{\textrm{max}}$}{
$\lambda_{\textrm{total}} = \frac{S(t) }{N-1} \texttt{sum}(\mathscr{V}^r) + \frac{S(t) }{N-1} \texttt{sum}(\mathscr{V}^i) + \eta R(t)$\;
$\tau_c \sim \textrm{Exp}(\lambda_{\textrm{total}})$\;
\eIf{$t + \tau_c < \textrm{min}(\texttt{peek}(\mathscr{U}^r),\texttt{peek}(\mathscr{U}^i))$}{
\tcp{Infection or waning comes first}
$u \sim \textrm{Uniform}(0,1)$\;
\eIf{$u < (\frac{S(t) }{N-1} \texttt{sum}(\mathscr{V}^r) + \frac{S(t) }{N-1} \texttt{sum}(\mathscr{V}^i)) / \lambda_{\textrm{total}}$}{
\tcp{Infection}
\eIf{$u < (\frac{S(t) }{N-1} \texttt{sum}(\mathscr{V}^r)) / \lambda_{\textrm{total}}$}{
$I_r(t + \tau_c) = I_r(t) + 1$\;
$S(t + \tau_c) = S(t) - 1$\;
$\texttt{push}(\mathscr{U}^r, t + \tau_c + 1)$\;
$\nu^r_{\textrm{new}} \sim \textrm{Gamma}(\gamma R_0^r/k,k)$\;
$\texttt{push}(\mathscr{V}^r, \nu^r_{\textrm{new}})$\;
}{
$I_i(t + \tau_c) = I_i(t) + 1$\;
$S(t + \tau_c) = S(t) - 1$\;
$\texttt{push}(\mathscr{U}^i, t + \tau_c + 1)$\;
$\nu^i_{\textrm{new}} \sim \textrm{Gamma}(\gamma R_0^i/k,k)$\;
$\texttt{push}(\mathscr{V}^i, \nu^i_{\textrm{new}})$\;
}
}{
\tcp{Waning}
$R(t + \tau_c) = R(t) - 1$\;
$S(t + \tau_c) = S(t) + 1$\;
}
$t = t + \tau_c$\;
}{
\tcp{Recovery comes first}
\eIf{$\texttt{peek}(\mathscr{U}^r) < \texttt{peek}(\mathscr{U}^i)$}{
$t_{\textrm{new}} = \texttt{pop}(\mathscr{U}^r)$\;
$\texttt{pop}(\mathscr{V}^r)$\;
$I_r(t_{\textrm{new}}) = I_r(t) - 1$\;
$R(t_{\textrm{new}}) = R(t) + 1$\;
$t = t_{\textrm{new}}$\;
}{
$t_{\textrm{new}} = \texttt{pop}(\mathscr{U}^i)$\;
$\texttt{pop}(\mathscr{V}^i)$\;
$I_i(t_{\textrm{new}}) = I_i(t) - 1$\;
$R(t_{\textrm{new}}) = R(t) + 1$\;
$t = t_{\textrm{new}}$\;
}
}

}

\caption{Algorithm to simulate the full epidemic process. Note that $\texttt{queue}()$ creates an empty queue data structure, $\texttt{push}(\mathscr{X},y)$ adds element $y$ to the end of queue $\mathscr{X}$, $\texttt{peek}(\mathscr{X})$ looks at the first value in $\mathscr{X}$ (without removing it), and  $\texttt{pop}(\mathscr{X})$ removes the first value in  $\mathscr{X}$, and returns it. For simplicity we do not include initialising the first $I_i$ individual in this description; it is introduced at time 1, with a transition from a susceptible individual, and its death time and individual reproductive capacity replace the first element in $\mathscr{U}^i$ and $\mathscr{V}^i$, respectively.}

\end{algorithm}

\subsection{Simulation study results}

\begin{figure*}
\begin{center}
  \includegraphics[width=0.95\textwidth]{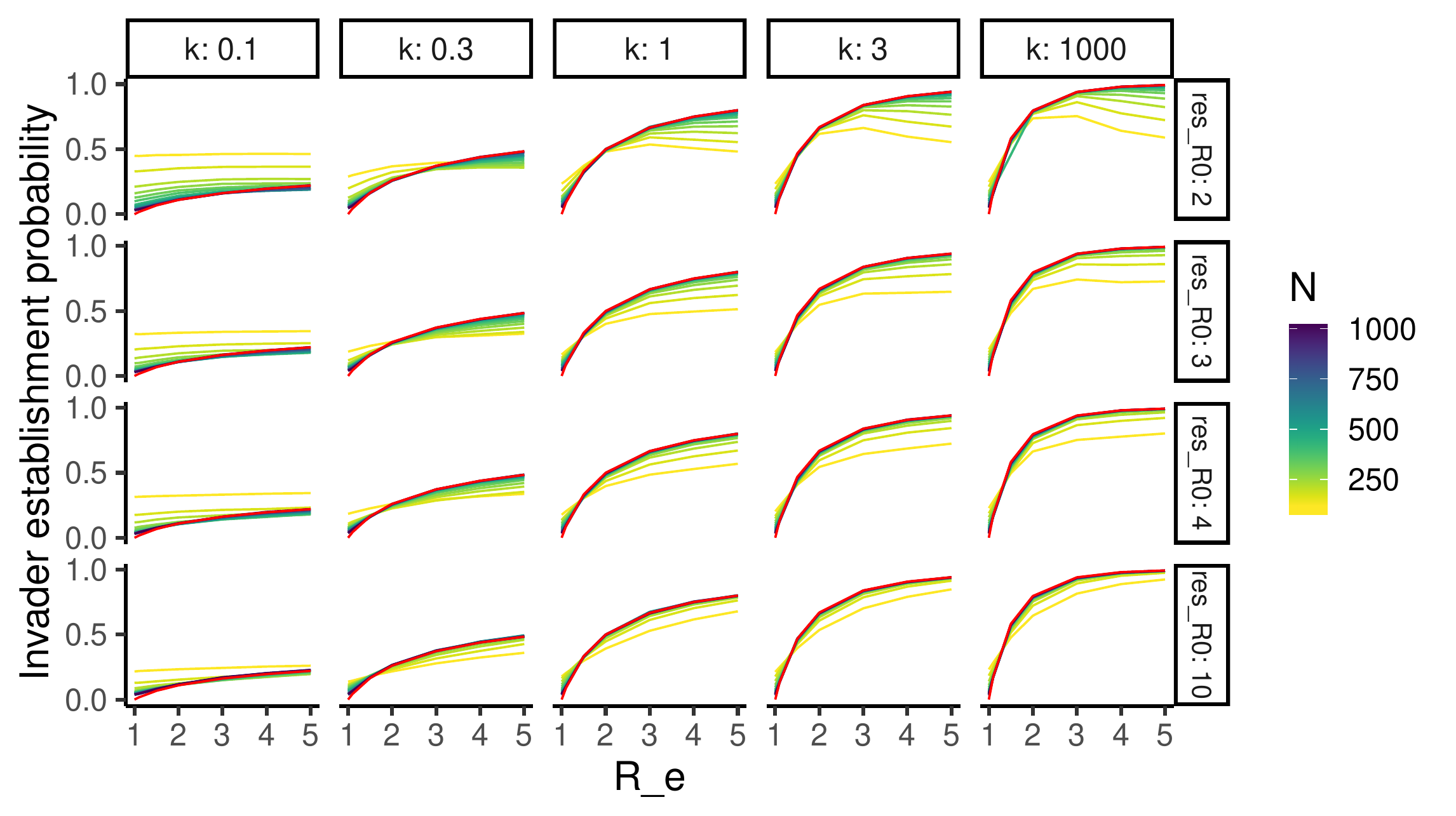}
\caption{Comparison between the branching process approximation for invader establishement success, and empirical invader establishment success in simulations of a full epidemic process. The process was simulated for a range of values of $k$, $R_0^r$, and $R_{\textrm{eff}}^i$ (i.e., $R_0^i / R_0^r$), as the total population size, $N$ varied from 100 to 1,000. The red line indicates the branching process approximation solution.}
\label{fig:simulations_grid}       
\end{center}
\end{figure*}

\begin{figure*}
\begin{center}
  \includegraphics[width=0.95\textwidth]{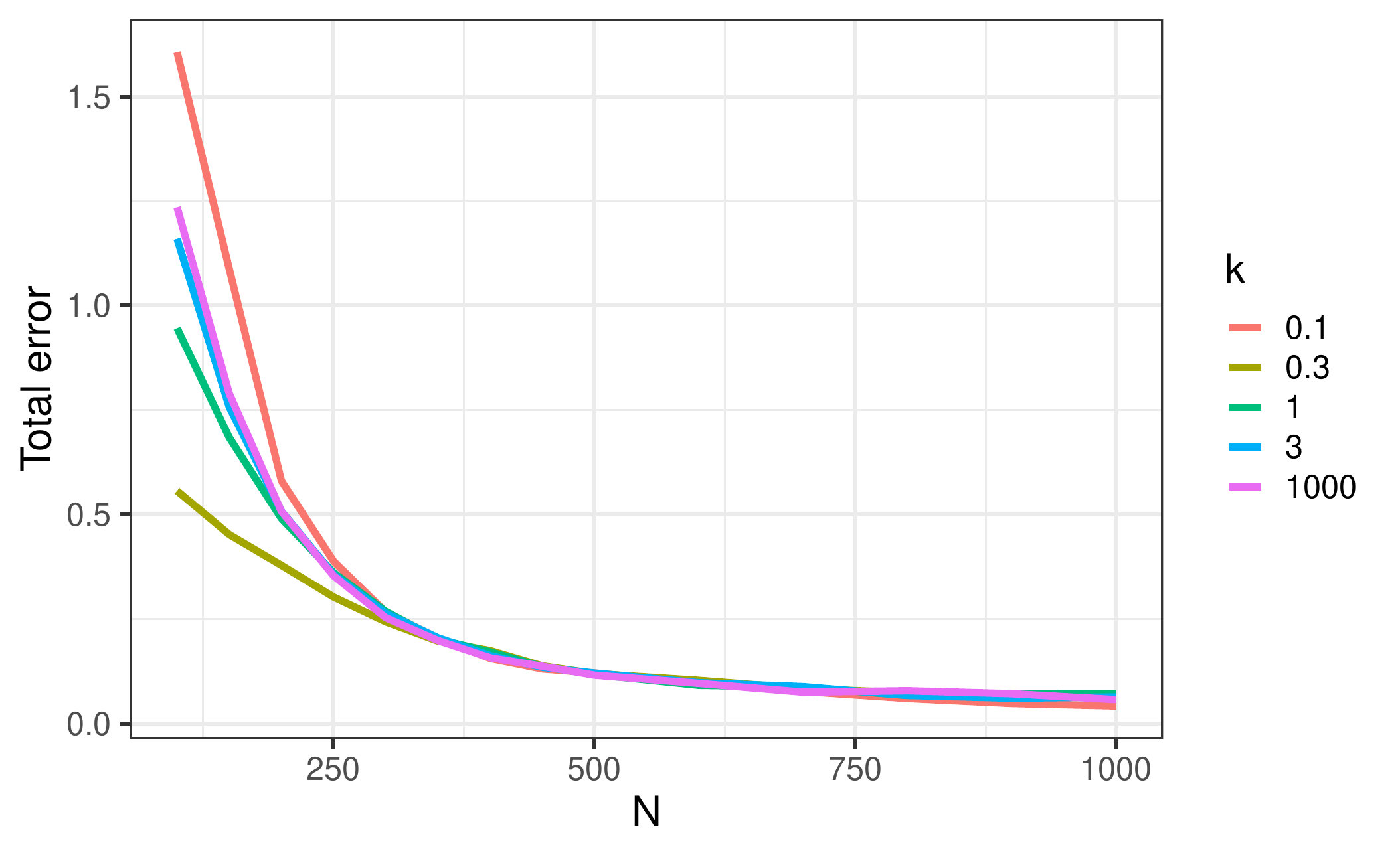}
\caption{Error between the branching process estimate of invader establishment probability, and establishment success in simulations of size $N$. $R_0^r$ is fixed at 2.0 in these simulations. Error is calculated as $\sum_{R_{\textrm{eff}} = 1}^5 \| \hat{q}_s(R_{\textrm{eff}}) - q(R_{\textrm{eff}}) \|$, where $\hat{q}_s(R_{\textrm{eff}})$ is the proportion of simulations in which establishment was successful at the given $R_{\textrm{eff}}$ level.}
\label{fig:total_error}       
\end{center}
\end{figure*}

\begin{figure*}
\begin{center}
  \includegraphics[width=0.95\textwidth]{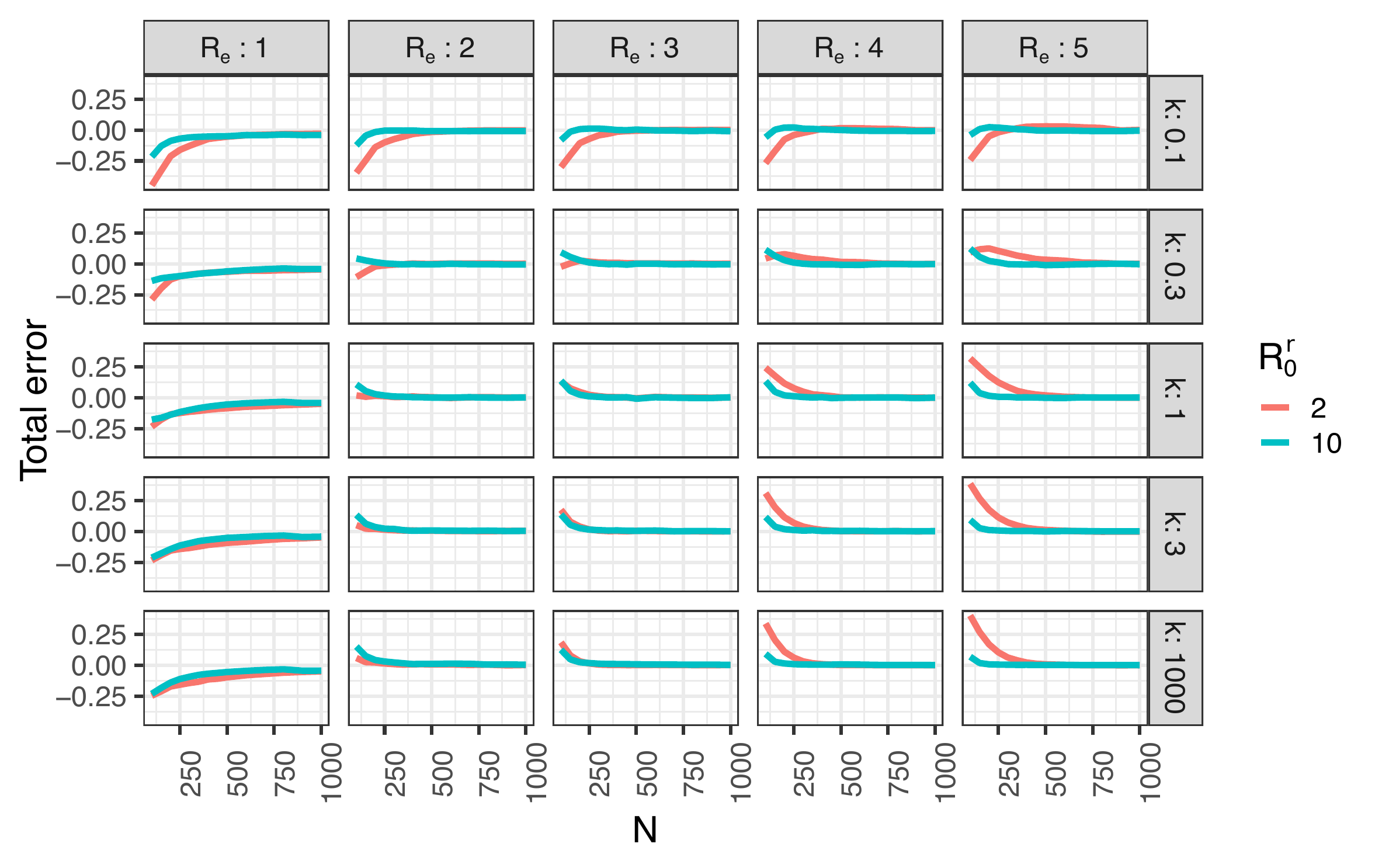}
\caption{The error: ($(1 - q) -$ proportion of sims in which the invader becomes established) at each individual parameter choice, by the total population size $N$. This shows that, when $R_{\textrm{eff}}^i$ is 1.0, or $k$ is 0.1, the branching process approximation underestimates the success of the invader; but when $R_{\textrm{eff}}^i$ is large, its success is overestimated, particularly when $R_0^r$ is small.}
\label{fig:grid_errors}       
\end{center}
\end{figure*}

When the simulated population size was small, there was substantial error between simulated establishment success of the invading strain, and the branching process approximation; however, as population size increased beyond 500, the error between these quantities reduced (Figure \ref{fig:total_error}). This behaviour was robust to variations in $k$, and to $R_0^r$ (Figure \ref{fig:simulations_grid}). 


The most substantial difference was that, particularly at smaller population sizes, the invader had some chance to become established even when $R_{\textrm{eff}}^i$ was 1.0; and this was more likely for more heterogeneous individual reproductive capacity distributions. Intuitively, this could occur when the initial invading individual was a `superspreader' (i.e., an individual with high individual reproductive capacity), but those individuals infected with the resident strain were not. 

\begin{figure*}
\begin{center}
  \includegraphics[width=0.95\textwidth]{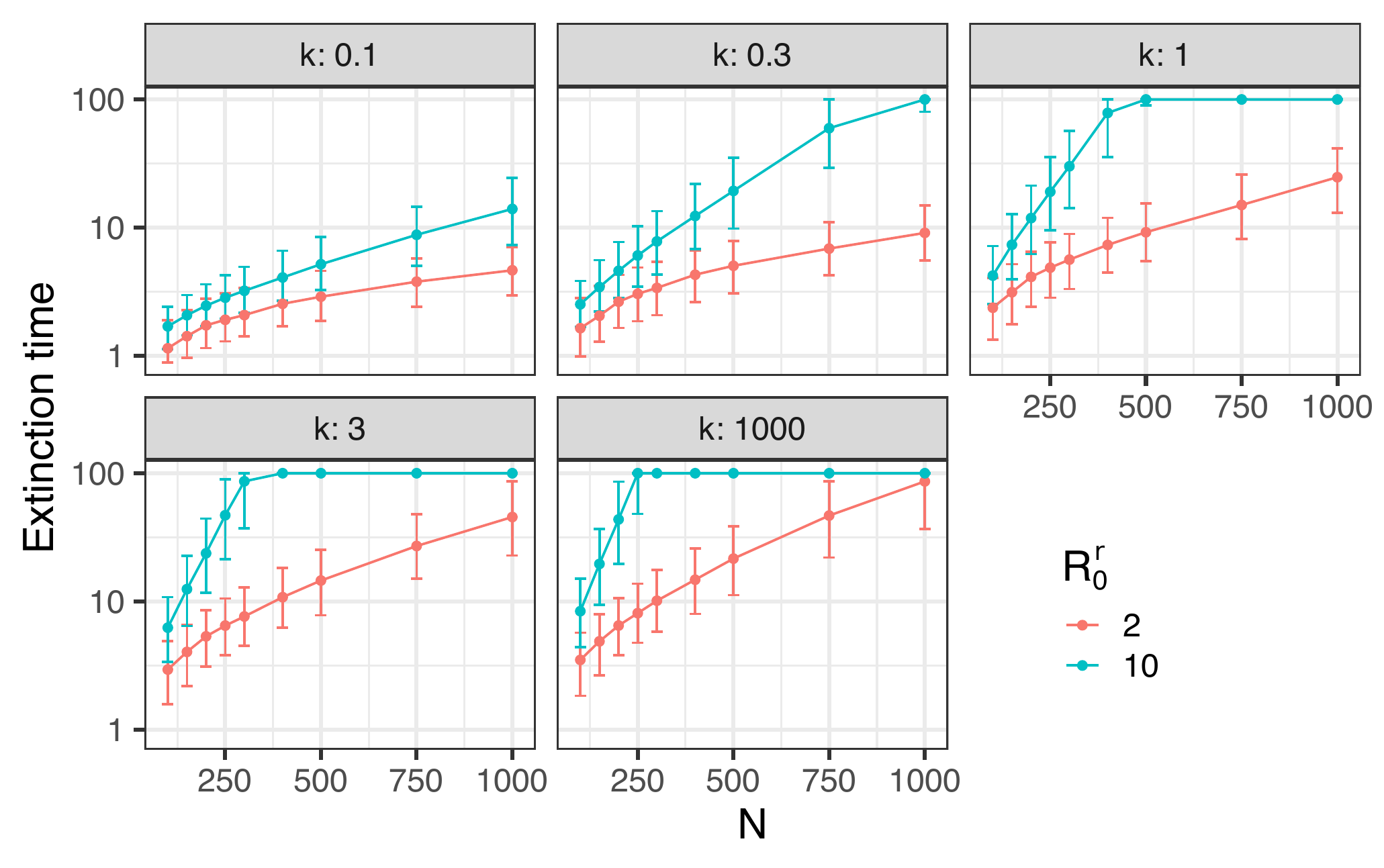}
\caption{Simulated extinction times for the endemic disease process (initiated in equilibrium at time 0), in the absence of the invader strain. Note that the simulation was truncated at time $t=100$, and so outbreaks that persisted at that time, were assigned that extinction time. Lines show median values, error bars indicate 25\% and 75\% quantiles.}
\label{fig:extTimes}       
\end{center}
\end{figure*}

More broadly, the interactions between $N$, $k$, the reproductive capacity of the resident strain, and the effective reproductive capacity of the invader have complex impacts on the correspondence between the branching process approximation and simulation results (Figure \ref{fig:total_error}). For intuition around the accuracy of the branching process approximation, we can consider two key drivers.
\begin{itemize}
    \item {\em The size of the population of susceptibles available to the invading strain, early in its outbreak.} A remark of Ball and Donnelly \cite{ball1995} suggests that, in the case they consider, the epidemic grows like a branching process until approximately $N^{1/2}$ of individuals in the population are infected. While the process we consider is not the same, we can use this as a heuristic. Rather than the full population, we are interested in the population of susceptible individuals, given the circulation of the endemic disease in the population. We denote this $S0$, and note that it is approximately $\frac{N}{R_0^r}$. We observe that for small values of $N$, the growth of the process can only be approximated by a branching process up to very small outbreaks, e.g., until 7 or fewer infected individuals when $N=100$.
    \item {\em The capacity of the resident strain (or the invader) to persist in the population.} In small populations, it is possible for the resident strain itself to become extinct quickly, due to the stochastic nature of disease transmission. As heterogeneity in individual reproductive capacity increases, this becomes more likely to occur, particularly at smaller values of $R_0$ (Figure \ref{fig:extTimes}). As such, when the invader is introduced, it becomes impossible to distinguish success of the invader strain from natural extinction of the resident; or failure of the invader from successful establishment followed quickly by its natural extinction.
\end{itemize}




In short, the branching process approximation is definitely to be avoided for small population sizes, but can produce reasonable results as population size increases (particularly under conditions when the endemic disease is capable of persisting for substantial time periods). We advise caution, and, where necessary, verifying results through simulation.

\end{document}